\documentclass[conference]{IEEEtran}

\usepackage{cite}
\usepackage{amsmath,amssymb,amsfonts}
\usepackage{algorithmic}
\usepackage{graphicx}
\usepackage{textcomp}
\usepackage{xcolor}
\usepackage[utf8]{inputenc}

\def\BibTeX{{\rm B\kern-.05em{\sc i\kern-.025em b}\kern-.08em
    T\kern-.1667em\lower.7ex\hbox{E}\kern-.125emX}}

\newcommand{\ddalphIn}{%
  \ensuremath{\Sigma_{\mathrm{In}}}%
}

\newcommand{\ddalphOut}{%
  \ensuremath{\Sigma_{\mathrm{Out}}}%
}

\newcommand{\ddcellWidth}{%
  2.95ex%
}
\newcommand{\ddcellHeight}{%
  2.25ex%
}
\newcommand{\ddcellN}[1]{%
  \makebox[\ddcellWidth]{\rule{0pt}{\ddcellHeight}\ensuremath{#1}}%
}
\newcommand{\ddcell}[1]{%
  \ddcellN{\mathtt{#1}}%
}
\newcommand{\ddpartHeight}{%
  5.5ex%
}
\newcommand{\ddpartSink}{%
  2.25ex%
}
\newcommand{\ddpart}[1]{%
  \makebox[\ddcellWidth]{%
    \rule[-\ddpartSink]{0pt}{\ddpartHeight}\ensuremath{#1}%
  }%
}

\newcommand{\ddHover}{%
  \ensuremath{\mathcal{O}_H}%
}
\newcommand{\ddVover}{%
  \ensuremath{\mathcal{O}_V}%
}

\newcommand{\ddextitlespace}{%
  3ex%
}

\begin{document}

\title{Efficient Encoding of Data\\ into Two-Dimensional Constrained
  Bit Patterns}

\author{
  \IEEEauthorblockN{Danny Dubé}
  \IEEEauthorblockA{\textit{Université Laval}\\ Quebec City, Canada
    \\ \texttt{Danny.Dube@ift.ulaval.ca}}
}

\maketitle

\begin{abstract}
  Two-dimensional constrained coding is a problem that is much more
  difficult than its one-dimensional counterpart.  Indeed, in two
  dimensions, obtaining the answers to very natural questions becomes
  uncomputable.  In particular, it is undecidable to determine if it
  is possible to fill the infinite plane with symbols in such a way
  that no forbidden pattern appears.  Also, even when we know that
  such an infinite plane exists, it is uncomputable to determine the
  maximal rate at which payload data can be embedded into the
  selection of a valid infinite plane.  Recently, Nakamura et
  al.~presented a technique that efficiently performs the construction
  of a matrix of symbols that embeds payload data.  Their technique is
  efficient in the sense that the construction takes time that is
  proportional to the area of the constructed matrix.  Their technique
  is based on the offline elaboration of a collection of tiles, which
  is then used for the matrix construction.  The
  collection-elaboration step is time consuming and it might even
  never terminate nor succeed.  In this work, we extend their
  technique by generalizing their notion of tile.  Our technique has
  the potential to achieve much higher data-embedding rates.
\end{abstract}


\section{Introduction}

\label{sect:intro}

\subsection{Constrained Coding}

Constrained coding consists in building strings on an
alphabet~\ddalphOut{} that obey a given constraint.  The constraint
that must be obeyed depends on the application.  For instance,
\emph{balanced codewords} are strings on $\ddalphOut = \{ \mathtt{0},
\mathtt{1} \}$ that contain exactly as many~'$\mathtt{0}$'s
as~'$\mathtt{1}$'s~\cite{knuth1986efficient}.  An application that
requires the use of balanced codes would fix a certain codeword
length~$M$, for $M$ even, and would use a coding function that would
map input data (or \emph{payload} data), which is a string of
arbitrary length on some input alphabet~\ddalphIn{}, into a number of
balanced codewords, which are each in~$\ddalphOut^M$.  For reasons of
efficiency, the mapping ought to be devised so that as few output
symbols get emitted per input symbol, on average, while obeying the
constraint of balance.  The balance constraint is said to be
\emph{global} because the property that must be obeyed applies on
whole codewords.

On the other hand, a constraint may be \emph{local}, in the sense that
any substring of length, say,~$l$ of a valid codeword has to obey a
certain criterion.  For instance, an application may require that any
run of a repeated symbol is limited in length.  Say, any run must have
maximal length~$4$.  At the same time, any run must also have minimal
length~$2$.  For technical reasons, it is often more convenient to
specify run-length limitations (RLL) in the NZRI notation.  In NZRI
notation, a zero indicates that the current symbol is identical to the
preceding symbol, while a one indicates that the current symbol
differs from the preceding symbol.  On a binary alphabet, there is a
one-to-one correspondence between the plain encoding of a string and
its NZRI encoding.  The NZRI encoding is especially convenient when we
are only interested in constraining the length of the runs, not the
actual symbols that constitute the runs.

Moreover, it is often more convenient to specify a local constraint by
listing the substrings that are \emph{forbidden}.  For instance, an
RLL code in the NZRI notation may be subject to a $(d, k)$-constraint,
which means that two successive ones must be separated from each other
by at least $d$~zeros but by no more than $k$~zeros.  In particular,
the $(1,3)$-RLL constraint states that there must at least one~($1$)
but at most three~($3$) zeros between two successive ones.  The
$(1,3)$-constraint is conveniently expressed by the two forbidden
patterns `$\mathtt{11}$' and `$\mathtt{0000}$'; in other words, the
first forbidden pattern expresses the fact that two ones cannot be
next to each other (i.e.~there cannot be as few as no zeros between
two successive ones) and four zeros cannot be consecutive (i.e.~there
cannot be four zeros or more between two successive ones).

\subsection{The Two-Dimensional Variant}

In this work, we consider the two-dimensional (2D) variant of
constrained coding.  In particular, we consider 2D constrained coding
with local constraints only.  A given instance of this problem can be
specified using a finite set of forbidden 2D finite patterns.
Forbidden patterns can have arbitrary shapes.  However, for the sake
of simplicity and without loss of generality, we may translate the set
of irregularly shaped forbidden patterns into an equivalent finite set
of rectangular \emph{valid} patterns; each of which is a $w \times h$
rectangle of symbols in~\ddalphOut~\cite{ota16}.  Then a matrix (in
the finite case) or a 2D plane (in the infinite case) on~\ddalphOut{}
is valid if, for any position, the $w \times h$ rectangle of symbols
that we extract at that position is valid.

It is well known that 2D constrained coding has to deal with
uncomputability.  In particular, it is generally undecidable to verify
if a given set of forbidden patterns allows the 2D infinite plane can
be filled in a valid way.  Also, even in cases in which we know that a
valid infinite plane exists, it is generally uncomputable to determine
the data-embedding capacity that can be achieved through the selection
of the symbol that appear in each position of the plane.

\subsection{The Task: Fast Data Encoding}

In this work, our goal consists in devising a fast procedure for
filling a matrix and for embedding data into the bits that get
selected.  We focus on the case $\ddalphOut = \{ \mathtt{0},
\mathtt{1} \}$.  It is important to note that we do not claim to bring
solutions on how to address the uncomputable aspects that hide behind
the task.  Our procedure relies on a preliminary step which is
performed offline and which addresses a difficult combinatorial task.
If the preliminary step completes with success, then we obtain the
means for performing matrix/plane filling.  It is just that the
preliminary step might never succeed.  The procedure that we propose
here is an extension of the work presented by Nakamura et
al.~\cite{nakamura18}.

\subsection{A Specific Coding Problem}

\begin{figure}
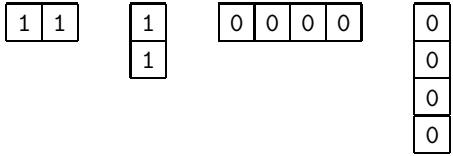

  \[
    \begin{array}{c@{\qquad}c@{\qquad}c@{\qquad}c}
        \begin{array}[t]{|@{}c@{}|@{}c@{}|}
        \hline
          \ddcell{1} & \ddcell{1}
        \\ \hline
        \end{array}
      &
        \begin{array}[t]{|@{}c@{}|}
        \hline
          \ddcell{1}
        \\ \hline
          \ddcell{1}
        \\ \hline
        \end{array}
      &
        \begin{array}[t]{|@{}c@{}|@{}c@{}|@{}c@{}|@{}c@{}|}
        \hline
          \ddcell{0} & \ddcell{0} & \ddcell{0} & \ddcell{0}
        \\ \hline
        \end{array}
      &
        \begin{array}[t]{|@{}c@{}|}
        \hline
          \ddcell{0}
        \\ \hline
          \ddcell{0}
        \\ \hline
          \ddcell{0}
        \\ \hline
          \ddcell{0}
        \\ \hline
        \end{array}
    \end{array}
  \]
  \caption{The forbidden patterns of 2D $(1,3)$-RLL constrained
    coding.}
  \label{fig:forbpats}
\end{figure}

In order to present our proposal and make a comparison possible with
the work by Nakamura et al., we address the same 2D constrained coding
problem that they addressed in their work.  The problem consists in
constructing a matrix or a plane while obeying the constraints which
are the straightforward 2D version of the usual 1D $(1,3)$-RLL
constraints.  More explicitly, we forbid two ones from occurring next
to each other, vertically or horizontally, and we forbid four zeros
from occurring consecutively, vertically or horizontally.
Figure~\ref{fig:forbpats} depicts the constraints that apply to the
addressed problem.

\section{The NOM Tile-Based Encoding Technique}

\label{sect:NOM}

We summarize the technique proposed by Nakamura et
al.~\cite{nakamura18}.  For the sake of brevity, we refer to their
technique as the \emph{NOM} tile-based encoding technique, after the
initials of the authors'~names.

\subsection{Tile-Based Encoding}

Basically, the NOM technique consists in a preliminary step which is
the offline elaboration of a collection of \emph{tiles} and then in
the regular data-encoding step.  A tile is a bit matrix of size $w
\times h$, for sufficiently large width~$w$ and height~$h$.  The
meaning of ``sufficiently large'' is explained below.  All the tiles
of the collection have to obey the constraints that are specific to 2D
$(1,3)$-RLL constrained coding.  Also, all the tiles of the collection
have the shape and contents shown in Figure~\ref{fig:NOMtiles}.  Note
that all the non-central parts of a tile are identical from tile to
tile; i.e.~the $\alpha$, $\beta$, and~$\gamma$ parts.  The central
part of a tile, however, varies and we say that the~$i$th~tile
has~$\delta_i$ as a central part.

\begin{figure}
  \[
    \begin{array}{|c|c|c|}
    \hline
      \ddpart{\alpha} & \ddpart{\beta}    & \ddpart{\alpha}
    \\ \hline
      \ddpart{\gamma} & \ddpart{\delta_i} & \ddpart{\gamma}
    \\ \hline
      \ddpart{\alpha} & \ddpart{\beta}    & \ddpart{\alpha}
    \\ \hline
    \end{array}
  \]
  \[
    \begin{array}{lll}
      \mbox{where} & \alpha : 3 \times 3, & \beta : (w-6) \times 3,
    \\
      & \gamma : 3 \times (h-6),\ \mbox{and} & \delta_i : (w-6) \times (h-6).
    \end{array}
  \]
  \caption{Shape and contents of the tiles in the NOM technique.}
  \label{fig:NOMtiles}
\end{figure}

The parts of the tiles that are constant ensure that, when tiling a
surface, either a matrix or the plane, that surface may be tiled
quickly, without ever requiring backtracking.  That is, if a tile is
placed at some position, then it is possible to place any tile of the
collection exactly $w-3$~columns to the right (or to the left) of it.
Similarly, it is possible to place any tile of the collection exactly
$h-3$~rows below (or above) it.  This tiling strategy causes a tile to
have an overlap of thickness~$3$ with its right-hand side neighbour,
with its left-hand side neighbour, with its neighbour above, and with
its neighbour below.  The contents of the overlaps are guaranteed to
agree, due to the parts that are constant.  Moreover, the constraints
of the specific constrained coding problem at hand are guaranteed to
be obeyed.  The thickness of~$3$ has been selected because it is the
minimal one that ensures the avoidance of forbidden patterns, given
that the largest forbidden patterns have size~$4$, horizontally or
vertically.  For example, if three consecutive zeros appear
horizontally in part~$\gamma$, then every part~$\delta_i$ is built in
such a way that it avoids extending the run of zeros up to length four
or more.

\subsection{Characteristics}

On top of providing constructive means to tile a surface, the NOM
technique allows one to embed information by the selection of the
tiles.  If the collection of tiles has size~$N$, then $\log_2 N$~bits
get embedded per tile.  Clearly, given that a collection of tiles has
been elaborated offline beforehand, then the tiling process and
information embedding can be performed with constant time per tile;
i.e.~in time proportional to the area of the covered surface.

The NOM technique has the advantage that all the costly (due to the
combinatorics) and risky (due to the semicomputability) computations
are performed offline.  It also has the advantage of defining a
fixed-to-fixed code.  Consequently, given the size of the data that
one wants embed into a constrained tiling, it is straightforward to
determine the area of a sufficiently large surface to hold that data.
On the other hand, a disadvantage of the technique comes from the
incapacity of the overlapping parts to encode data.  Indeed, every
$h-3$~rows, there are $3$~rows that do not encode any payload data.
Vertically too, every $w-3$~columns, there are $3$~columns that no
encode any payload data.  Another disadvantage is that, if one wants
to obtain a reasonable data embedding capacity, relatively large tiles
have to be used; i.e.~the area of~$\delta_i$ has to compare favourably
enough to those of~$\alpha$, $\beta$, and~$\gamma$.

\section{Proposed Tile-Based Encoding Technique}

\label{sect:prop}

In this work, we propose an alternate technique that alleviates this
problem and that has the potential to embed more information per coded
bit.  We start by presenting the proposed technique, then we discuss
about its characteristics, and we end the section with a concrete
example.

\subsection{Tile-Based Encoding}

Like the NOM technique, ours is based on overlapping tiles but it
differs in imposing a looser constraint on the overlapping parts.
Instead of forcing the overlap between a tile and its neighbour to the
right be always the same, we allow multiple different overlaps.  We
relax the constraint on the overlap between vertical neighbours in a
similar fashion.

\begin{figure}
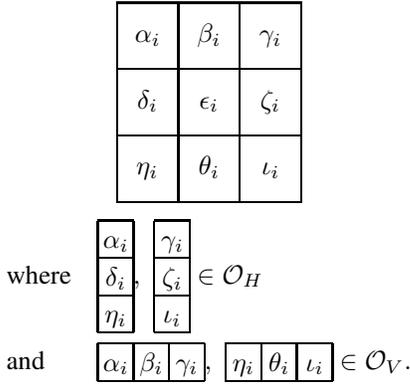

  \[
    \begin{array}{|c|c|c|}
    \hline
      \ddpart{\alpha_i} & \ddpart{\beta_i}    & \ddpart{\gamma_i}
    \\ \hline
      \ddpart{\delta_i} & \ddpart{\epsilon_i} & \ddpart{\zeta_i}
    \\ \hline
      \ddpart{\eta_i}   & \ddpart{\theta_i}   & \ddpart{\iota_i}
    \\ \hline
    \end{array}
  \]
  \[
    \begin{array}{ll}
        \mbox{where}
      &
          \begin{array}{|@{}c@{}|}
          \hline
            \ddcellN{\alpha_i}
          \\ \hline
            \ddcellN{\delta_i}
          \\ \hline
            \ddcellN{\eta_i}
          \\ \hline
          \end{array}
          ,\ 
          \begin{array}{|@{}c@{}|}
          \hline
            \ddcellN{\gamma_i}
          \\ \hline
            \ddcellN{\zeta_i}
          \\ \hline
            \ddcellN{\iota_i}
          \\ \hline
          \end{array}
        \in
          \ddHover
    \\
        \mbox{and}\rule{0pt}{3ex}
      &
          \begin{array}{|@{}c@{}|@{}c@{}|@{}c@{}|}
          \hline
              \ddcellN{\alpha_i}
            &
              \ddcellN{\beta_i}
            &
              \ddcellN{\gamma_i}
          \\ \hline
          \end{array}
          ,\ 
          \begin{array}{|@{}c@{}|@{}c@{}|@{}c@{}|}
          \hline
              \ddcellN{\eta_i}
            &
              \ddcellN{\theta_i}
            &
              \ddcellN{\iota_i}
          \\ \hline
          \end{array}
        \in
          \ddVover.
    \end{array}
  \]
  \caption{Shape and contents of the tiles in our proposed technique.}
  \label{fig:proptiles}
\end{figure}

\begin{figure}
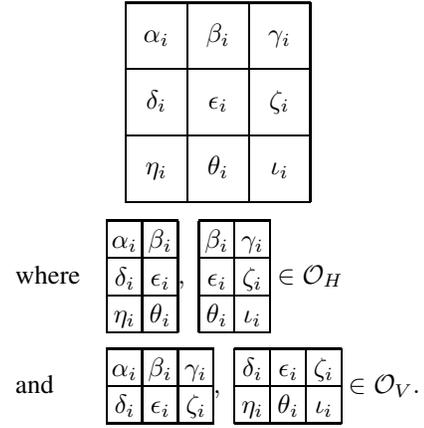

  \[
    \begin{array}{|c|c|c|}
    \hline
      \ddpart{\alpha_i} & \ddpart{\beta_i}    & \ddpart{\gamma_i}
    \\ \hline
      \ddpart{\delta_i} & \ddpart{\epsilon_i} & \ddpart{\zeta_i}
    \\ \hline
      \ddpart{\eta_i}   & \ddpart{\theta_i}   & \ddpart{\iota_i}
    \\ \hline
    \end{array}
  \]
  \[
    \begin{array}{ll}
        \mbox{where}
      &
          \begin{array}{|@{}c@{}|@{}c@{}|}
          \hline
            \ddcellN{\alpha_i} & \ddcellN{\beta_i}
          \\ \hline
            \ddcellN{\delta_i} & \ddcellN{\epsilon_i}
          \\ \hline
            \ddcellN{\eta_i}   & \ddcellN{\theta_i}
          \\ \hline
          \end{array}
          ,\ 
          \begin{array}{|@{}c@{}|@{}c@{}|}
          \hline
            \ddcellN{\beta_i}    & \ddcellN{\gamma_i}
          \\ \hline
            \ddcellN{\epsilon_i} & \ddcellN{\zeta_i}
          \\ \hline
            \ddcellN{\theta_i}   & \ddcellN{\iota_i}
          \\ \hline
          \end{array}
        \in
          \ddHover
    \\
        \mbox{and}\rule{0pt}{5ex}
      &
          \begin{array}{|@{}c@{}|@{}c@{}|@{}c@{}|}
          \hline
              \ddcellN{\alpha_i}
            &
              \ddcellN{\beta_i}
            &
              \ddcellN{\gamma_i}
          \\ \hline
              \ddcellN{\delta_i}
            &
              \ddcellN{\epsilon_i}
            &
              \ddcellN{\zeta_i}
          \\ \hline
          \end{array}
          ,\ 
          \begin{array}{|@{}c@{}|@{}c@{}|@{}c@{}|}
          \hline
              \ddcellN{\delta_i}
            &
              \ddcellN{\epsilon_i}
            &
              \ddcellN{\zeta_i}
          \\ \hline
              \ddcellN{\eta_i}
            &
              \ddcellN{\theta_i}
            &
              \ddcellN{\iota_i}
          \\ \hline
          \end{array}
        \in
          \ddVover.
    \end{array}
  \]
  \caption{Shape and contents of narrow and shallow tiles in our
    proposed technique.}
  \label{fig:smalltiles}
\end{figure}

Let us suppose that we use tiles of size $w \times h$.  Then, the
elaboration of a collection of tiles starts with the selection of a
set of valid horizontal overlaps~\ddHover{} and a set of valid
vertical overlaps~\ddVover{}.  A valid overlap in~\ddHover{} is a
matrix of size~$3 \times h$ that does not contain a forbidden pattern.
A valid overlap in~\ddVover{} is a matrix of size~$w \times 3$ that
does not contain a forbidden pattern.  The invariants that we impose
on a valid collection of tiles, given~\ddHover{} and~\ddVover{}, are
the following:
\begin{itemize}
\item[\textbf{C1}] for any tile in the collection, the sub-matrix
  obtained by extracting its leftmost $3$~columns and the one obtained
  by extracting its rightmost $3$~columns are elements of~\ddHover{};
  moreover, the sub-matrix obtained by extracting its top $3$~rows and
  the one obtained by extracting its bottom $3$~rows are elements
  of~\ddVover{} and
\item[\textbf{C2}] given any leftmost overlap~$t_L$ taken
  from~\ddHover{} and any top overlap~$t_T$ taken from~\ddVover{}, if
  the top $3 \times 3$ sub-matrix of~$t_L$ is equal to the leftmost $3
  \times 3$ sub-matrix of~$t_T$, then there exists at least one tile
  in the collection that has its $3$~leftmost columns equal to~$t_L$
  and its $3$~top rows equal to~$t_T$.
\end{itemize}
These invariants ensure that, if we tile a surface from left to right
and downwards, then it is always possible to pick a tile in the
collection to cover the next tile position.
Figure~\ref{fig:proptiles} depicts the parts that constitute a tile of
the collection.  Note that our technique does not require tiles to be
large; i.e.~such that we have~$w > 6$ and~$h > 6$.  In that respect,
Figure~\ref{fig:proptiles} is misleading in that it implies that the
leftmost overlap is disjoint from the rightmost overlap; similarly
regarding the top and the bottom overlaps.
Figure~\ref{fig:smalltiles} depicts the shape and contents of the
tiles for the case~$w \leq 6$ and~$h \leq 6$.  Note that there are, in
principle, two other cases to describe (that is, the case~$w > 6$
and~$h \leq 6$ and the case~$w \leq 6$ and~$h > 6$) but we trust that
the reader can infer the invariants that apply in these cases from
those that are already depicted in Figures~\ref{fig:proptiles}
and~\ref{fig:smalltiles}.

\subsection{Characteristics}

The technique we propose has more numerous similarities with the NOM
technique than differences.  It is based on elaborating a collection
of tiles offline that may then be used to cover a surface very
quickly.  Selecting a tile for the next tile position can be made in
constant time.  The overlap between neighbouring tiles also has
thickness~$3$.  However, our technique allows the contents of the
overlap to vary from a tile position to another.  In principle, this
allows our technique to have the overlaps embed payload data as well
as the central parts~$\epsilon_i$.  All the costly and risky
computations are performed during the offline step only.

One disadvantage of our technique is that it is not necessarily a
fixed-to-fixed code like the NOM technique.  Indeed, each time a tile
must be selected subject to the given overlaps~$t_L$ and~$t_T$, there
may be a different number of candidates available.  This is made
obvious in Subsection~\ref{sect:example}.  This means that the
selection process that is performed at each tile position may lead to
a different number of bits being embedded.  Another disadvantage of
our technique, and it is a more severe one, is that
condition~\textbf{C2} of the invariants only guarantee that there is
at least \emph{one} candidate available.  This condition is sufficient
to ensure progress in covering an arbitrarily large surface.  However,
it is not sufficient to guarantee that an arbitrary number of bits
will get embedded in the process.  Although the consequences of that
weakness may look bad, this is easy to fix.  The corrective is
mentioned as future work, in the conclusion; see
Section~\ref{sect:conclusion}.

Finally, an advantage that our technique brings is that, since all the
parts of the tiles have the potential to embed payload data, it is not
necessary to use as large tiles as in the NOM technique.  Indeed, in
the NOM technique, it is crucial to use tiles of size at least~$7
\times 7$, to ensure that central part~$\delta_i$ is non-trivial.  In
our technique, it makes sense to consider using tiles of size~$6
\times 6$ or smaller.  This is made obvious in the example below.

\subsection{Example of a Collection of $4 \times 4$ Tiles}

\label{sect:example}

We have implemented a prototype for the proposed technique in order to
obtain a proof of concept.  Although we expected that smaller tiles
than in the NOM technique would likely be sufficient, it came as a
surprise to us that a collection of tiles of the smallest possible
size could be elaborated; i.e.~a collection of $4 \times 4$ tiles.
Let us note that such small tiles allow no more than a single bit to
be selected per selected tile, as all the bits in a new tile except
the bottom-right one are already determined by the overlaps~$o_L$
and~$o_T$; see Figure~\ref{fig:freebits4x4}(d).

\begin{figure}
  \[
    \begin{array}{|@{}c@{}|@{}c@{}|@{}c@{}|@{}c@{}|}
    \hline
      \ddcell{?} & \ddcell{?} & \ddcell{?} & \ddcell{?}
    \\ \hline
      \ddcell{?} & \ddcell{?} & \ddcell{?} & \ddcell{?}
    \\ \hline
      \ddcell{?} & \ddcell{?} & \ddcell{?} & \ddcell{?}
    \\ \hline
      \ddcell{?} & \ddcell{?} & \ddcell{?} & \ddcell{?}
    \\ \hline
      \multicolumn{4}{@{}c@{}}{
        \makebox[0pt]{(a) Top-left}\rule{0pt}{\ddextitlespace}
      }
    \end{array}
    \quad
    \begin{array}{|@{}c@{}|@{}c@{}|@{}c@{}|@{}c@{}|}
    \hline
      \ddcell{0} & \ddcell{1} & \ddcell{0} & \ddcell{?}
    \\ \hline
      \ddcell{1} & \ddcell{0} & \ddcell{0} & \ddcell{?}
    \\ \hline
      \ddcell{0} & \ddcell{0} & \ddcell{1} & \ddcell{?}
    \\ \hline
      \ddcell{1} & \ddcell{0} & \ddcell{0} & \ddcell{?}
    \\ \hline
      \multicolumn{4}{@{}c@{}}{
        \makebox[0pt]{(b) Top row}\rule{0pt}{\ddextitlespace}
      }
    \end{array}
    \quad
    \begin{array}{|@{}c@{}|@{}c@{}|@{}c@{}|@{}c@{}|}
    \hline
      \ddcell{0} & \ddcell{1} & \ddcell{0} & \ddcell{1}
    \\ \hline
      \ddcell{1} & \ddcell{0} & \ddcell{0} & \ddcell{0}
    \\ \hline
      \ddcell{0} & \ddcell{0} & \ddcell{1} & \ddcell{0}
    \\ \hline
      \ddcell{?} & \ddcell{?} & \ddcell{?} & \ddcell{?}
    \\ \hline
      \multicolumn{4}{@{}c@{}}{
        \makebox[0pt]{(c) Left col.}\rule{0pt}{\ddextitlespace}
      }
    \end{array}
    \quad
    \begin{array}{|@{}c@{}|@{}c@{}|@{}c@{}|@{}c@{}|}
    \hline
      \ddcell{0} & \ddcell{1} & \ddcell{0} & \ddcell{1}
    \\ \hline
      \ddcell{1} & \ddcell{0} & \ddcell{0} & \ddcell{0}
    \\ \hline
      \ddcell{0} & \ddcell{0} & \ddcell{1} & \ddcell{0}
    \\ \hline
      \ddcell{1} & \ddcell{0} & \ddcell{0} & \ddcell{?}
    \\ \hline
      \multicolumn{4}{@{}c@{}}{
        \makebox[0pt]{(d) Other pos.}\rule{0pt}{\ddextitlespace}
      }
    \end{array}
  \]
  \caption{Potentially free bits in tiles at different position on the
    surface to cover.}
  \label{fig:freebits4x4}
\end{figure}

Out of the set of the $236$~valid $4 \times 4$~tiles, a collection
made of $213$~tiles which obeys~\textbf{C1} and~\textbf{C2} could be
elaborated by our prototype.  Due to lack of space, we do not present
the collection of tiles itself.  We rather mention a few facts about
the collection.  As written above, in most tile positions, only one
bit may potentially be selected (see Figure~\ref{fig:freebits4x4}(d)),
which means that the constraints imposed by the forbidden patterns are
quite strict.  Indeed, out of the $195$~possible contexts set by the
overlaps~$t_L$ and~$t_T$, only~$18$ allow the bottom-right bit to be
chosen freely.  It means that the $177$~other contexts cause the
bottom-right bit to be either a zero or a one by
obligation.\footnote{From the $18$~contexts that allow freedom and the
  $177$~ones that force the choice of the bit, we obtain the $2 * 18 +
  177 = 213$ tiles of our collection.}

\begin{figure}
  \begin{center}
    \fbox{\resizebox{0.9\columnwidth}{!}{\includegraphics{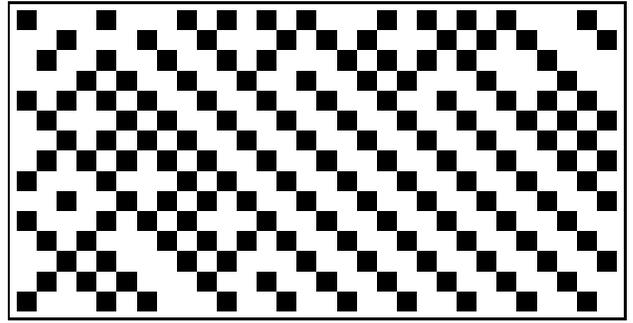}}}
  \end{center}
  \caption{Binary matrix obtained by embedding ``\texttt{Hello
      world!}''  into 2D $(1,3)$-RLL constrained code using $4 \times
    4$ tiles.  A black square represents a bit set to one.}
  \label{fig:hellomatrix}
\end{figure}

We tested the data-embedding capacity of our collection of tiles on a
short string of payload bits.  The data are made from the
concatenation of the $7$-bit \textsc{Ascii} codewords for the
characters of the text ``\texttt{Hello world!}''.  The data has been
successfully embedded into a $30 \times 15$~binary matrix.
Figure~\ref{fig:hellomatrix} shows the resulting matrix.

In order to get an idea of where data get embedded into the binary
matrix, Figure~\ref{fig:nbchoices} shows, for each tile position, how
many tiles were candidates to be selected in that position.  Let us
recall that there are fewer tile positions than bits in the binary
matrix; this is because the tiles have size~$4 \times 4$.  We can
observe that there are usually more options available along the
left-hand and top boundaries and even more so at the top-left
position.  This is natural as the selection of a tile in the top-left
position is made in the context shown in
Figure~\ref{fig:freebits4x4}(a); in the top row, it is made in a
context like the one shown in Figure~\ref{fig:freebits4x4}(b); in the
leftmost column, it is made in a context like the one shown in
Figure~\ref{fig:freebits4x4}(c); and, elsewhere in the matrix, it is
made in a context like the one shown in
Figure~\ref{fig:freebits4x4}(d).  Clearly, very few bits of payload
data get embedded in the ordinary tile positions (``elsewhere'') on
average; most tile positions allow only a single candidate.  This
illustrates a point made above about the fact that our definition of a
valid collection does not guarantee that an arbitrary amount of
payload data can get embedded in arbitrary large binary matrices.  In
theory, it could happen that an unfortunate choice of tiles on a row
would constrain the following rows to deal only with contexts that
force a single-candidate tile selection.  This theoretically possible
phenomenon would not compromise the progress in tiling an arbitrarily
large surface but it could stop the progress in embedding payload
data.

\begin{figure*}
  \[
  \begin{array}{%
      |@{}c@{}|@{}c@{}|@{}c@{}|@{}c@{}|@{}c@{}|@{}c@{}|@{}c@{}%
      |@{}c@{}|@{}c@{}|@{}c@{}|@{}c@{}|@{}c@{}|@{}c@{}|@{}c@{}%
      |@{}c@{}|@{}c@{}|@{}c@{}|@{}c@{}|@{}c@{}|@{}c@{}|@{}c@{}%
      |@{}c@{}|@{}c@{}|@{}c@{}|@{}c@{}|@{}c@{}|@{}c@{}|@{}c@{}%
      |@{}c@{}|@{}c@{}|%
    }
  \hline
      \ddcellN{\resizebox{0.8\width}{\height}{\ensuremath{213}}}
    &
      \ddcellN{2} & \ddcellN{3} & \ddcellN{1} & \ddcellN{4}
    &
      \ddcellN{1} & \ddcellN{2} & \ddcellN{4} & \ddcellN{2}
    &
      \ddcellN{4} & \ddcellN{2} & \ddcellN{4} & \ddcellN{2} & \ddcellN{2}
    &
      \ddcellN{3} & \ddcellN{2} & \ddcellN{3} & \ddcellN{4}
    &
      \ddcellN{2} & \ddcellN{4} & \ddcellN{2} & \ddcellN{3} & \ddcellN{4}
    &
      \ddcellN{2} & \ddcellN{3} & \ddcellN{2} & \ddcellN{2}
  \\ \hline
      \ddcellN{2} & \ddcellN{2} & \ddcellN{1} & \ddcellN{1} & \ddcellN{1}
    &
      \ddcellN{1} & \ddcellN{1} & \ddcellN{1} & \ddcellN{1}
    &
      \ddcellN{1} & \ddcellN{1} & \ddcellN{1} & \ddcellN{2} & \ddcellN{1}
    &
      \ddcellN{1} & \ddcellN{1} & \ddcellN{1} & \ddcellN{2}
    &
      \ddcellN{1} & \ddcellN{1} & \ddcellN{1} & \ddcellN{1} & \ddcellN{1}
    &
      \ddcellN{2} & \ddcellN{1} & \ddcellN{1} & \ddcellN{1}
  \\ \hline
      \ddcellN{3} & \ddcellN{1} & \ddcellN{1} & \ddcellN{1} & \ddcellN{1}
    &
      \ddcellN{1} & \ddcellN{1} & \ddcellN{1} & \ddcellN{1}
    &
      \ddcellN{1} & \ddcellN{1} & \ddcellN{1} & \ddcellN{1} & \ddcellN{1}
    &
      \ddcellN{1} & \ddcellN{1} & \ddcellN{1} & \ddcellN{1}
    &
      \ddcellN{1} & \ddcellN{2} & \ddcellN{1} & \ddcellN{1} & \ddcellN{1}
    &
      \ddcellN{1} & \ddcellN{2} & \ddcellN{1} & \ddcellN{1}
  \\ \hline
      \ddcellN{4} & \ddcellN{2} & \ddcellN{1} & \ddcellN{2} & \ddcellN{1}
    &
      \ddcellN{1} & \ddcellN{1} & \ddcellN{1} & \ddcellN{1}
    &
      \ddcellN{1} & \ddcellN{1} & \ddcellN{1} & \ddcellN{1} & \ddcellN{1}
    &
      \ddcellN{1} & \ddcellN{1} & \ddcellN{1} & \ddcellN{1}
    &
      \ddcellN{1} & \ddcellN{1} & \ddcellN{1} & \ddcellN{1} & \ddcellN{1}
    &
      \ddcellN{1} & \ddcellN{1} & \ddcellN{2} & \ddcellN{1}
  \\ \hline
      \ddcellN{3} & \ddcellN{1} & \ddcellN{2} & \ddcellN{1} & \ddcellN{2}
    &
      \ddcellN{1} & \ddcellN{1} & \ddcellN{1} & \ddcellN{1}
    &
      \ddcellN{1} & \ddcellN{1} & \ddcellN{1} & \ddcellN{1} & \ddcellN{1}
    &
      \ddcellN{1} & \ddcellN{1} & \ddcellN{1} & \ddcellN{1}
    &
      \ddcellN{1} & \ddcellN{1} & \ddcellN{1} & \ddcellN{1} & \ddcellN{1}
    &
      \ddcellN{1} & \ddcellN{1} & \ddcellN{1} & \ddcellN{2}
  \\ \hline
      \ddcellN{2} & \ddcellN{1} & \ddcellN{1} & \ddcellN{2} & \ddcellN{1}
    &
      \ddcellN{1} & \ddcellN{1} & \ddcellN{1} & \ddcellN{1}
    &
      \ddcellN{1} & \ddcellN{1} & \ddcellN{1} & \ddcellN{1} & \ddcellN{1}
    &
      \ddcellN{1} & \ddcellN{1} & \ddcellN{1} & \ddcellN{1}
    &
      \ddcellN{1} & \ddcellN{1} & \ddcellN{1} & \ddcellN{1} & \ddcellN{1}
    &
      \ddcellN{1} & \ddcellN{1} & \ddcellN{1} & \ddcellN{1}
  \\ \hline
      \ddcellN{4} & \ddcellN{1} & \ddcellN{1} & \ddcellN{1} & \ddcellN{2}
    &
      \ddcellN{1} & \ddcellN{2} & \ddcellN{1} & \ddcellN{1}
    &
      \ddcellN{1} & \ddcellN{1} & \ddcellN{1} & \ddcellN{1} & \ddcellN{1}
    &
      \ddcellN{1} & \ddcellN{1} & \ddcellN{1} & \ddcellN{1}
    &
      \ddcellN{1} & \ddcellN{1} & \ddcellN{1} & \ddcellN{1} & \ddcellN{1}
    &
      \ddcellN{1} & \ddcellN{1} & \ddcellN{1} & \ddcellN{1}
  \\ \hline
      \ddcellN{4} & \ddcellN{1} & \ddcellN{1} & \ddcellN{1} & \ddcellN{1}
    &
      \ddcellN{2} & \ddcellN{1} & \ddcellN{2} & \ddcellN{1}
    &
      \ddcellN{1} & \ddcellN{1} & \ddcellN{1} & \ddcellN{1} & \ddcellN{1}
    &
      \ddcellN{1} & \ddcellN{1} & \ddcellN{1} & \ddcellN{1}
    &
      \ddcellN{1} & \ddcellN{1} & \ddcellN{1} & \ddcellN{1} & \ddcellN{1}
    &
      \ddcellN{1} & \ddcellN{1} & \ddcellN{1} & \ddcellN{1}
  \\ \hline
      \ddcellN{1} & \ddcellN{1} & \ddcellN{2} & \ddcellN{1} & \ddcellN{1}
    &
      \ddcellN{1} & \ddcellN{2} & \ddcellN{1} & \ddcellN{2}
    &
      \ddcellN{1} & \ddcellN{1} & \ddcellN{1} & \ddcellN{1} & \ddcellN{1}
    &
      \ddcellN{1} & \ddcellN{1} & \ddcellN{1} & \ddcellN{1}
    &
      \ddcellN{1} & \ddcellN{1} & \ddcellN{1} & \ddcellN{1} & \ddcellN{1}
    &
      \ddcellN{1} & \ddcellN{1} & \ddcellN{1} & \ddcellN{1}
  \\ \hline
      \ddcellN{3} & \ddcellN{2} & \ddcellN{1} & \ddcellN{2} & \ddcellN{1}
    &
      \ddcellN{1} & \ddcellN{1} & \ddcellN{1} & \ddcellN{1}
    &
      \ddcellN{2} & \ddcellN{1} & \ddcellN{1} & \ddcellN{1} & \ddcellN{1}
    &
      \ddcellN{1} & \ddcellN{1} & \ddcellN{1} & \ddcellN{1}
    &
      \ddcellN{1} & \ddcellN{1} & \ddcellN{1} & \ddcellN{1} & \ddcellN{1}
    &
      \ddcellN{1} & \ddcellN{1} & \ddcellN{1} & \ddcellN{1}
  \\ \hline
      \ddcellN{4} & \ddcellN{1} & \ddcellN{1} & \ddcellN{1} & \ddcellN{2}
    &
      \ddcellN{1} & \ddcellN{1} & \ddcellN{1} & \ddcellN{2}
    &
      \ddcellN{1} & \ddcellN{1} & \ddcellN{1} & \ddcellN{1} & \ddcellN{1}
    &
      \ddcellN{1} & \ddcellN{1} & \ddcellN{1} & \ddcellN{1}
    &
      \ddcellN{1} & \ddcellN{1} & \ddcellN{1} & \ddcellN{1} & \ddcellN{1}
    &
      \ddcellN{1} & \ddcellN{1} & \ddcellN{1} & \ddcellN{1}
  \\ \hline
      \ddcellN{2} & \ddcellN{1} & \ddcellN{1} & \ddcellN{1} & \ddcellN{1}
    &
      \ddcellN{2} & \ddcellN{1} & \ddcellN{1} & \ddcellN{1}
    &
      \ddcellN{1} & \ddcellN{2} & \ddcellN{1} & \ddcellN{1} & \ddcellN{1}
    &
      \ddcellN{1} & \ddcellN{1} & \ddcellN{1} & \ddcellN{1}
    &
      \ddcellN{1} & \ddcellN{1} & \ddcellN{1} & \ddcellN{1} & \ddcellN{1}
    &
      \ddcellN{1} & \ddcellN{1} & \ddcellN{1} & \ddcellN{1}
  \\ \hline
    \end{array}
  \]
  \caption{Number of choices available in the selection for each tile
    position.}
  \label{fig:nbchoices}
\end{figure*}

This study of the same 2D constrained coding problem should be
extended to larger tiles, like tiles of sizes~$5 \times 5$, $6 \times
6$,~\ldots\ \ However, as a proof of concept, it is already
interesting to see that tiles as small as~$4 \times 4$ manage to be
sufficient to embed possibly unlimited amounts of payload data into
binary matrices.  The use of larger tiles would render
single-candidate selections much less frequent, if not impossible.
Future work presented in the conclusion addresses this issue.

Studying our collection of $4 \times 4$~tiles further might lead to
the establishment of stronger results about the data-embedding
capacity of the whole matrix or, to the very least, its leftmost and
top portions.  In particular, the leftmost column and the top row are
simpler to study because their processes of tile selection are
one-dimensional Markov processes that can be modelled using finite
automata.  A finite automaton can be scrutinized to determine whether
the control could get stuck into a subset of states where no data
embedding occurs.  If such an undesirable possibility exists, then the
removal of certain transitions may turn the automaton into one where
no such traps exist.  One the other hand, formally studying the
tile-selection process elsewhere in the matrix might prove difficult,
if computable at all.  Indeed, a finite automaton is not sufficient to
model that process as the tile-selection process in not only
influenced by the overlaps left by the left-hand side neighbours
during the tiling of a row but also by the sequence of top overlaps
left by the tiles selected in the previous row.  Collectively, these
top overlaps form a memory with potentially arbitrarily complex
states.

\section{Conclusion}

\label{sect:conclusion}

In this work, we presented an extension of a tile-based technique for
efficiently encoding data into 2D binary matrices while obeying 2D
constraints on the bit patterns.  Both the technique by Nakamura et
al.~and ours proceed by elaborating a collection of tiles offline and
then using that collection to tile an arbitrarily large surface in
time proportional to the area of the surface.  Ours differs from the
previous one in the fact that our tiles do not need to contain fixed
parts for ensuring coordination between neighbouring tile.  This
feature causes our technique to have the potential to achieve much
higher data embedding rates.  We presented the definitions that are
necessary for our technique and we made a case study, based on our
prototype.

Investigation ought to be continued on our technique.  First, we
intend to use stronger conditions on the collections that we elaborate
in order to guarantee not only that the tiling process may continue
indefinitely but also that additional payload data can get embedded
indefinitely by the tiling process.  This added guarantee can be
obtained by further requiring that, in any context, the selection of a
tile is made over \emph{at least two} candidates.

\newpage

Second, when multiple candidate tiles may fit a particular context, it
might be more profitable not to use equal probabilities over the
candidates.  Indeed, even if equiprobable selection maximizes the
immediate entropy during that very selection, it does not necessarily
maximizes entropy on the long run.  Favouring a particular tile may
reduce the (expected) number of steps before another
multiple-candidate tile selection is reached.  Clearly, the
probabilities associated with the different candidates ought to be
adjusted in order to optimize the global data-embedding rate.  Due to
the 2D nature of the tile-selection process, the optimization might
prove to be a difficult task.

\section*{Acknowledgements}

The author wishes to thank the authors of the paper on the NOM
technique~\cite{nakamura18} for the invaluable discussions about their
work.  Also, this research and meetings with the authors of the paper
on the NOM technique were greatly facilitated by the Invitational
Fellowship for Research in Japan granted by the Japanese Society for
the Promotion of Science.

\bibliographystyle{IEEEtran}
\bibliography{article}

\nocite{louidor08}

\end{document}